\documentclass[journal,twoside]{IEEEtran}
\usepackage{color}
\usepackage{cite}
\ifCLASSINFOpdf
\usepackage[pdftex]{graphicx}
\else
\fi
\usepackage[cmex10]{amsmath}
\usepackage{amssymb}
\usepackage{amsfonts}
\usepackage{float}
\usepackage{subfigure}
\usepackage{multirow}
\usepackage[capitalize]{cleveref}
\usepackage{algorithm}
\usepackage{algpseudocode}
\usepackage{multirow}
\usepackage{multicol}
\usepackage{todonotes}
\usepackage{bm}

\usepackage{amsmath}

\DeclareMathOperator*{\argmin}{arg\,min}

\usepackage{tikz}
\tikzstyle{cnode} = [draw, circle,scale=.6]
\tikzstyle{level 1} = [level distance=.35\textwidth, sibling distance=1\textwidth]
\tikzstyle{level 2} = [level distance=.35\textwidth, sibling distance=.45\textwidth]
\tikzstyle{level 3} = [level distance=.3\textwidth, sibling distance=.25\textwidth]
\usetikzlibrary{decorations.pathreplacing,patterns,external}

\usetikzlibrary{external}
\ifCLASSOPTIONcompsoc
\usepackage[caption=false,font=normalsize,labelfont=sf,textfont=sf]{subfig}
\else
\usepackage[caption=false,font=footnotesize]{subfig}
\fi

\newcommand{\rev}[1]{{\color{black}{#1}}}
\newcommand{\revnew}[1]{{\color{black}{#1}}}
\newcommand{\ylrev}[1]{{\color{black}{#1}}}
\begin{document}
	\title{Detecting Resonance of Radio-Frequency Cavities Using Fast Direct Integral Equation Solvers and Augmented Bayesian Optimization}
	\author{Yang Liu, \textit{Member}, \textit{IEEE}, Tianhuan Luo, Aman Rani, Hengrui Luo, and Xiaoye Sherry Li
		\thanks{Manuscript received XX XX, XXXX. (\textit{Corresponding author: Yang Liu.})}
		
		\thanks{Y. Liu, H. Luo, and X. S. Li are with Applied Mathematics and Computational Research Division, Lawrence Berkeley National Laboratory, Berkeley, CA 94720, USA. (e-mail: liuyangzhuan@lbl.gov, hrluo@lbl.gov, XSLi@lbl.gov). }
		
		\thanks{T. Luo is with Accelerator Technology and Applied Physics Division, Lawrence Berkeley National Laboratory, Berkeley, CA 94720, USA. (e-mail: TLuo@lbl.gov). }

		\thanks{A. Rani is with the department of Mathematics and Statistics, Texas Tech University, Lubbock, TX, 79409, USA. (e-mail: aman.rani@ttu.edu).}
	}
	\markboth{IEEE JOURNAL ON MULTISCALE AND MULTIPHYSICS COMPUTATIONAL TECHNIQUES,~Vol.~P, No.~PP, Month~Year}
	{LIU \MakeLowercase{\textit{et al.}}: Computing Resonance Cavities}
	
	\maketitle
	
	\begin{abstract}
		This paper presents a computationally efficient framework for identifying resonance modes of 3D radio-frequency (RF) cavities with damping waveguide ports. The proposed framework relies on surface integral equation (IE) formulations to convert the task of resonance detection to the task of finding resonance frequencies at which the lowest few eigenvalues of the system matrix is close to zero. For the linear eigenvalue problem \rev{with a fixed frequency}, we propose leveraging fast direct solvers to efficiently invert the system matrix; for the frequency search problem, we develop a hybrid optimization algorithm that combines Bayesian optimization with down-hill simplex optimization. The proposed IE-based resonance detection framework (IERD) has been applied to detection of high-order resonance modes (HOMs) of realistic accelerator RF cavities to demonstrate its efficiency and accuracy. 
	\end{abstract}
	

	%
	\IEEEpeerreviewmaketitle

	\section{Introduction}\label{sec:intro}
	%
	%
	%
	%

	\IEEEPARstart{H}{igh}-quality radio-frequency (RF) cavities are critical components for high energy particle accelerators developed at national facilities and laboratories \cite{qiang2019integration,luo2019rf,stanek2019overview}. RF cavities are usually \revnew{excited by a time-harmonic source field operated at the working resonance frequency to support a particular electromagnetic field pattern (called the fundamental resonance mode)}, which provides stable acceleration of the particle bunches passing through the cavity. However, the transient bunch, which can be treated as e.g., a Gaussian wavelet packet in time, can also excite undesired high-order resonance modes (HOMs) \revnew{at higher resonance frequencies} which can cause deterioration of beam quality and severe damage to the device. These HOMs are often suppressed by adding damping waveguide ports to the cavity, which couple the HOMs from the cavity to the external absorbing loads. \rev{Recall that the quality (Q) factor of a cavity is defined as the ratio of stored energy in the cavity and the dissipated power per cycle by the cavity wall losses and/or damping ports. While the fundamental mode operates with high Q factors, the HOMs requires sufficiently low Q factors so that they cannot be established in practice.}   
	
	Given a fixed design with cavity geometry, material, port position and port shapes, the fundamental mode and HOMs \revnew{and their resonance frequencies} need to be accurately and efficiently modeled by numerical tools. The most commonly-used numerical method for realistic 3D cavity modeling is the finite element method (FEM) \cite{jin2015finite,xiao2019advances,lee2009omega3p}, which formulates the problem as an eigenvalue problem whose eigenvalues and eigenvectors are respectively the resonance frequencies and resonant electrical fields in the cavity volume \revnew{(or called eigenmodes). In this paper, we use the terms ``eigenmode'' and ``resonance mode'' interchangeably.} However, FEM faces the following computational challenges: (1) The FEM method requires spatial discretization of the entire cavity volume with dense meshes and high-order basis functions to resolve the resonance modes, which lead to large numbers of unknowns and high CPU/memory usage. (2) When waveguide boundary conditions are used for the damping ports or the dissipative power loss is considered, the FEM method leads to a nonlinear eigenvalue problem, and linearization techniques such as Newton's method \cite{liao2010nonlinear,voss2004arnoldi}, contour integral methods \cite{beyn2012integral,asakura2009numerical}, and rational Krylov methods \cite{van2018computing}. However, these methods suffer from either lack of convergence \rev{and/or} high computational costs. Moreover, the FEM method tends to become less accurate when the resonance frequency lies near or below the lowest cutoff frequencies of the ports. (3) Although the Q factors can be computed using the FEM method, they cannot provide information about accuracy of the computed resonance frequencies regarding uncertainty in design parameters and numerical models.     
	
	This paper considers the surface integral equation (IE) method as an alternative to the FEM method to address the abovementioned computational challenges. The IE method formulates the problem as searching for the frequencies at which the system matrix's smallest few eigenvalues are close to zero. In contrast to FEM, the IE method only requires spatial discretization of the cavity boundary leading to much smaller numbers of unknowns. In addition, at each trial frequency, a linear dense eigenvalue problem needs to be solved and traditional linear eigen solvers can be directly applied. Last but not least, the IE method naturally provides bandwidth information about the computed resonance frequencies \ylrev{by investigating a sequence of trial frequencies and the associated eigenvalues}. That being said, the IE methods have been primarily considered only for traveling wave modeling \cite{chang2015higher,wang2015higher,Auda1983junction,bunger2000moment} and characteristic mode analysis (CMA) \cite{chen2015characteristic,huang2019unified,harrington1971theory}, and the use of IE methods for resonance modeling remains largely unexplored \cite{angiulli2020computation,huang2018study}. This is due to the following two computational difficulties: (1) Unlike traveling wave modeling and CMA, the resonance problem leads to intrinsically large-scale and ill-conditioned linear systems at resonance frequencies of the HOMs. Although fast iterative algorithms such as fast multipole methods \cite{chew2001fast} have been applied to large-scale CMA \cite{dai2016large}, they become computationally infeasible for resonance problems. \ylrev{This is due to the ill-conditioned IE matrix near the resonance frequencies which, even with good preconditioners, lead to large numbers of matrix-vector multiplications.} (2) The identification of the resonance frequencies \ylrev{is} typically performed with grid search or frequency sweep, which becomes computationally very expensive with fine frequency steps \rev{\cite{angiulli2020computation}}. Optimization algorithms \rev{that aim at minimizing the eigenvalues corresponding to each mode} can be considered instead of grid search, but the objective function w.r.t frequency exhibits sharp minimum for high-Q modes and flattened minimum for low-Q ones, and existing optimization tools \rev{either require too many samples to capture all resonance modes or suffer from convergence issues} for these complex objective functions. Therefore IE methods that can both robustly solve the eigenvalue problems with given trial frequencies and efficiently propose promising trial frequencies are called for.        
	
	The proposed \rev{IE-based resonance detection framework} (IERD) framework addresses the abovementioned two challenges by leveraging fast direct IE solvers and developing hybrid optimization algorithms. First, we rely on the shift-and-invert implicitly restarted Arnoldi method (IRAM) \cite{lehoucq1998arpack} for solving the eigenvalue problem for the smallest few eigenvalues, which requires supplying the inverse of the dense IE matrix. The matrix inverse is computed with fast direct solvers such as hierarchically off-diagonal low-rank (HODLR) matrices \cite{ambikasaran2013mathcal} and hierarchically off-diagonal butterfly (HODBF) \cite{liu2016hss}, among the more general class of hierarchical matrix algorithms \cite{Corona2015ON,hackbusch2003introduction,vandebril2005bibliography,Chris2019Adaptive,ma2019direct}. For working modes and the first few HOMs, the cavity remains electrically small and low-rank-based direct solvers, e.g., HODLR, are very efficient; for HOMs with higher frequencies, the cavity becomes electrically large.  Therefore, the equation system show certain patterns for which  butterfly-based direct solvers, e.g., HODBF becomes computational more efficient. In fact, butterfly-based direct solvers have been successfully applied to surface IE \cite{Han_2017_butterflyLUPEC,liu2016hss}, volume IE \cite{sayed2022VIE}, and differential equation \cite{liu2020MF}-based large-scale electromagnetic simulations. The use of HODLR and HODBF direct solvers can significantly improve the robustness and efficiency for solving the eigenvalue problem with fixed frequencies. Next, we propose a hybrid frequency search algorithm that combines the global search algorithm such as Gaussian process (GP)-based Bayesian optimization (BO) \cite{williams2006gaussian} and local search algorithm such as the down-hill simplex method (also known as the Nelder–Mead method) \cite{nelder1965simplex}. Our proposed algorithm first uses surrogate-based Bayesian optimization to locate the resonance regions in the frequency band of interest and then uses down-hill simplex algorithm to refine the optimal  frequency within region. Both steps require only smaller numbers of trial frequencies compared with brute-force frequency sweep or grid-search 
	scheme.            
	
	The rest of this paper is organized as follows. Section \ref{sec:form} formulates the resonance detection problem as an IE-based eigenvalue problem with trial frequencies. The fast direct solvers for accelerating the eigenvalue problem for each frequency is summarized in Section \ref{sec:directIE} and the proposed hybrid Bayesian optimization and down-hill simplex algorithm is detailed in Section \ref{sec:BO}. Section \ref{sec:result} presents a few canonical and real-life numerical examples to demonstrate the efficiency, accuracy and robustness of the proposed IERD framework.     
	
	\begin{figure}[!t]
		\centering
		\includegraphics[width=\columnwidth]{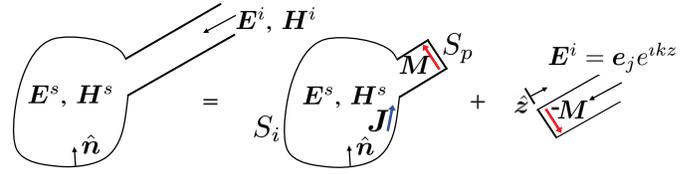}
		\caption{	
			Illustration of a RF cavity example with a waveguide port, decomposed as an interior problem and an exterior problem. It is assumed that the incident field $\bm{E}^i=\bm{e}_je^{\imath kz}$ has a single waveguide mode. \label{fig:cavity}}
	\end{figure}
	\section{Formulation}\label{sec:form}
	Consider the example of a RF cavity with a semi-infinite waveguide port in Fig. \ref{fig:cavity}. In practice, the port serves as a damping port with a finite length and absorbing materials at the end. For illustration purposes, we initially consider a single port in this section. Later, in Section \ref{sec:result}, we demonstrate the applicability of IERD to cases with multiple ports by providing numerical results. The cavity is filled with material of permittivity $\varepsilon_0$,  permeability $\mu_0$, and intrinsic impedance ${\eta_0}=\sqrt{\mu_0/\varepsilon_0}$. To modify the cavity geometry and material properties for a specific problem, adjust the values of $\varepsilon_0$, $\mu_0$, and $\eta_0$ accordingly. Additionally, you can update the cavity and waveguide port geometries by changing their dimensions and shapes in the mathematical formulation.
	In order to derive a formulation for the resonance modes, it is assumed that the cavity is excited by an incident field $\{\bm{E}^i(\bm{r}), \bm{H}^i(\bm{r})\}$ of frequency $f$ (or wavenumber $k$) through the port. The incident field excites surface equivalent source that produces a scattered field $\{\bm{E}^s(\bm{r}), \bm{H}^s(\bm{r})\}$ in the cavity. 
	
	The problem can be treated as the superposition of an interior problem and an exterior problem. The interior problem consists of a closed cavity with surface $S=S_i+S_p$, \revnew{where $S_i$ corresponds to the open surface of the original open cavity, and the added boundary $S_p$ corresponds to the surface of the port that encloses $S_i$.} 
	The scattered field $\{\bm{E}^s(\bm{r}), \bm{H}^s(\bm{r})\}$ is produced by a surface electric current density $\bm{J}(\bm{r})$ for $\bm{r}\in S$ and a surface magnetic current density $\bm{M}(\bm{r})$ for $\bm{r}\in S_p$ (\revnew{see Fig. \ref{fig:cavity} middle}). The exterior problem consists of the semi-infinite wavguide with the incident electric field $\bm{E}^i=\bm{e}_je^{\imath kz}$, where $z$ is the local coordinate with $z=0$ at $S_p$, and $\bm{e}_j$ is the normalized eigenvector of the waveguide (\revnew{see Fig. \ref{fig:cavity} right}). $\bm{e}_j$ can be TM or TE modes, and can have an analytical expression for rectangular or circular waveguides, or can be numerically computed for irregular-shaped waveguides. The scattered field $\{\bm{E}^s(\bm{r}), \bm{H}^s(\bm{r})\}$ can be reproduced by introducing an additional surface magnetic current density $-\bm{M}$ at $S_p$. 
	
	Enforcing the boundary conditions for electric fields on $S_i$ and $S_p$ and magnetic fields on $S_p$ leads to the following equations:
	\begin{align}
	&\hat{\bm{n}}\times\hat{\bm{n}}\times\mathcal{T}[\bm{J}](\bm{r}) - \hat{\bm{n}}\times\mathcal{K}[\frac{\bm{M}}{\eta_0}](\bm{r})-\frac{Z_s}{\eta_0}\bm{J}(\bm{r})=0, \bm{r}\in S_i\label{eq:Ei}	\\
	&\hat{\bm{n}}\times\hat{\bm{n}}\times\mathcal{T}[\bm{J}](\bm{r}) - \hat{\bm{n}}\times\mathcal{K}[\frac{\bm{M}}{\eta_0}](\bm{r})+\hat{\bm{n}}\times\frac{\bm{M}(\bm{r})}{\eta_0}=0, \bm{r}\in S_p\label{eq:Ep} \\
	&\frac{1}{\eta_0}\hat{\bm{n}}\times\bm{J}(\bm{r}) + \frac{1}{2}\sum_i\frac{\hat{\bm{z}}\times \bm{e}_i(\bm{r})}{\eta_i\eta_0}\int_{S_p}d\bm{r}'(\hat{\bm{z}}\times \bm{e}_i(\bm{r}'))\cdot \bm{M}(\bm{r}')\nonumber \\
	&=\frac{\hat{\bm{z}}\times \bm{e}_j(\bm{r})}{\eta_j\eta_0}, \bm{r}\in S_p \label{eq:Hp}
	\end{align}
	
	Here, $\hat{\bm{n}}$ denotes the inward surface normal, $\hat{\bm{z}}$ denotes the unit vector along the waveguide. $Z_s=(1+\imath)\sqrt{\frac{\pi f\mu_0}{\sigma_s}}$ is the surface impedance of the cavity wall material. Here $\sigma_s$ is the electric conductivity of the material and we use $5.80\times10^7$ S/m of Copper in this paper. $\eta_i$ and $\bm{e}_i(\bm{r})$, $i=1,2,\ldots$, are the impedance and normalized tangential electric field of waveguide mode $i$ with cutoff frequency $f_{c,i}<f$. Note that for regular-shaped wave guide ports such as circular or rectangular waveguides, $\bm{e}_i(\bm{r})$ are analytically defined; otherwise they can be numerically computed and tabulated from e.g., FEM methods. In (\ref{eq:operator}), the integral operators $\mathcal{T}$ and $\mathcal{K}$ are:
	\begin{align}
	\mathcal{T}[{\bm{X}}]({\bm{r}})&=&\imath{k}\int_S d\bm{r}' {\bm{X}}({\bm{r}'})\cdot \left( \bm{I}-\frac{1}{k^2}\nabla\nabla'\cdot
	\right)  {\bm{g}}(\bm{r},\bm{r}')\\
	\mathcal{K}[{\bm{X}}]({\bm{r}})&=&\hat{\bm{n}}\times\int_S d\bm{r}'{\bm{X}}({\bm{r}'})\times\nabla'{\bm{g}}(\bm{r},\bm{r}')
	\label{eq:operator}
	\end{align} 
	To numerically detect the resonance modes, $\bm{J}$ is discretized with $N_J$ local basis functions $\bm{f}_n(\bm{r})$, $n=1,\ldots,N_J$ such as the Rao-Wilton-Glisson (RWG) functions \cite{rao1982electromagnetic} or the high-order ($p$th order) Graglia-Wilton-Peterson (GWP($p$)) basis functions \cite{graglia1997higher} (RWG is essentially GWP(0)), $\bm{M}/\eta_0$ is discretized with $N_M$ mode basis functions $\hat{\bm{z}}\times\bm{e}_n$, $n=1,\ldots,N_M$ with $N_M$ denotes the number of propagating modes of the port whose cutoff frequencies $f_{c,n}<f$. 
	\begin{eqnarray}
	\bm{J}(\bm{r})=\sum_{n=1}^{N_J}I^J_n\bm{f}_n(\bm{r})\\
	\frac{\bm{M}(\bm{r})}{\eta_0}=\sum_{n=1}^{N_M}I^M_n\hat{\bm{z}}\times\bm{e}_n(\bm{r})
	\end{eqnarray}
	
	Next we test (\ref{eq:Ei}) and (\ref{eq:Ep}) with $\bm{f}_n$, (\ref{eq:Hp}) with $\hat{\bm{z}}\times\bm{e}_n$ and set the right-hand side (RHS) of (\ref{eq:Hp}) to 0. \rev{Note if $f$ is a resonance frequency in the band of interest $[f_{\min},f_{\max}]$, the system will require no excitation of the RHS}. The following eigenvalue system is constructed:
	\begin{eqnarray} 
	\begin{bmatrix}
	\bar{\bar{\mathbf{T}}} & \bar{\bar{\mathbf{K}}}\\
	\bar{\bar{\mathbf{C}}} & \bar{\bar{\mathbf{D}}}
	\end{bmatrix}\begin{bmatrix}\bar{\mathbf{I}}^J \\ \bar{\mathbf{I}}^M\end{bmatrix}\approx\bar{\mathbf{0}}.\label{eq:IEmatrix}
	\end{eqnarray}
	Note that ``$\approx$'' instead of ``$=$'' is used to account for numerical errors in surface discretization and matrix evaluation. In (\ref{eq:IEmatrix}), $\bar{\mathbf{I}}^J$ and $\bar{\mathbf{I}}^M$ are collections of $I^J_n$ and $I^M_n$, respectively. The $(m,n)^{th}$ entry of the sub-matrices \ylrev{is}:
	\begin{align}
	T_{mn} &= \langle \bm{f}_m(\bm{r}),\hat{\bm{n}}\times\hat{\bm{n}}\times\mathcal{T}[\bm{f}_n](\bm{r}) \rangle - \frac{Z_s}{\eta_0}\langle \bm{f}_m(\bm{r}),\delta^i(\bm{r})\bm{f}_n(\bm{r}) \rangle \\
	K_{m,n} &= \langle \bm{f}_m(\bm{r}),-\hat{\bm{n}}\times\mathcal{K}[\hat{\bm{z}}\times\bm{e}_n](\bm{r}) \rangle + \langle \bm{f}_m(\bm{r}),\hat{\bm{z}}\times\bm{e}_n(\bm{r}) \rangle \\
	C_{m,n} &= \langle\hat{\bm{z}}\times\bm{e}_m(\bm{r}),\frac{1}{\eta_0}\hat{\bm{n}}\times\bm{f}_n(\bm{r})\rangle \\
	D_{m,n} &= \frac{\delta_{m,n}}{2\eta_m}.
	\end{align}  
	Here, $\delta^i(\bm{r})=1$ if $\bm{r}\in S_i$ and 0 elsewhere, $\delta_{m,n}$ is the Kronecker delta function, and $\langle\cdot,\cdot\rangle$ denotes standard inner product. The system in (\ref{eq:IEmatrix}) can be written as $\bar{\bar{\mathbf{Z}}}(f)\bar{\mathbf{I}}\approx\bar{\mathbf{0}}$ with $\bar{\mathbf{I}}=[\bar{\mathbf{I}^J};\bar{\mathbf{I}}^M]$ and $\bar{\bar{\mathbf{Z}}}=[\bar{\bar{\mathbf{T}}},\bar{\bar{\mathbf{K}}};\bar{\bar{\mathbf{C}}},\bar{\bar{\mathbf{D}}}]$. 
	Therefore, the problem of resonance detection reduces to finding the trial frequency $f$ at which the smallest few eigenvalues of $\bar{\bar{\mathbf{Z}}}$ are close to 0. \ylrev{The physical meaning of such a formulation is that at resonance, the system can sustain an approximately non-decaying surface electrical current density without any external excitation (i.e., setting the RHS of the linear system to $\bar{\mathbf{0}}$).}
	
	The proposed IERD framework \rev{aims at the accurately and efficiently finding all resonance modes in the frequency band of interest.} IERD relies on a fast direct IE solver-based linear eigensolver to compute these eigenvalues for each trial frequency, as described in Section \ref{sec:directIE}, and an efficient optimization algorithm to search for the resonance frequencies of multiple modes, as detailed in Section \ref{sec:BO}.

	\section{Fast Direct Method-based Eigensolver}\label{sec:directIE}
	For a fixed frequency $f$, the proposed IERD framework utilizes the shift-and-invert IRAM method from ARPACK \cite{lehoucq1998arpack} with zero shift, which requires the input of the inverse operator $\bar{\bar{\mathbf{Z}}}^{-1}$. \ylrev{IRAM is a fast iterative algorithm that can be used to quickly compute the first few eigenvalues of interest. When the number of required eigenvalues is small, IRAM typically converges in a small number of iterations.} ARPACK returns the smallest few eigenvalues and their eigenvectors among which some eigenvectors do not correspond to a physical resonance, but rather artifact from numerical discretization errors. \ylrev{These eigenvectors are characterized as sparse vectors with few large elements and small ones elsewhere, representing non-physical and ``local" resonance near shape geometry regions. The sparsity of the eigenvectors can be measured by $||\bar{\mathbf{I}}||_1$ given that $||\bar{\mathbf{I}}||_\infty=1$.} Hence only those eigenvectors $\bar{\mathbf{I}}$ with $||\bar{\mathbf{I}}||_1/||\bar{\mathbf{I}}||_\infty>\epsilon_{norm}$ for a prescribed threshold $\epsilon_{norm}$ (e.g., $\epsilon_{norm}=1000$) are considered to be physically meaningful resonance vectors. \revnew{In this paper, we use ``eigenmodes'' to denote these physically meaningful resonance vectors. Note that those eigenvectors with $||\bar{\mathbf{I}}||_1/||\bar{\mathbf{I}}||_\infty<\epsilon_{norm}$ cannot be called eigenmodes as they are due to numerical modeling errors.} One frequency $f$ may yield no eigenmode or multiple eigenmodes. \revnew{Ideally assuming no material loss, power damping, numerical error and degenerate eigenmodes, each eigenmode has only one resonance frequency (i.e., $\bar{\bar{\mathbf{Z}}}$ is singular at the resonance frequency and the Q factor becomes infinity). In practice, however, a cavity can operate at near-resonance regime (i.e., slightly perturbed resonance frequency) with the same eigenmode. Therefore, one trial frequency $f$ can detect multiple eigenmodes when it's in the near-resonance regime of multiple eigenmodes.} 
	
	HODLR and HODBF are two types of rank-structured fast direct solvers for fast computing $\bar{\bar{\mathbf{Z}}}^{-1}$ implemented in the software ButterflyPACK \cite{osti_1564244}. These solvers first reorder the matrix $\bar{\bar{\mathbf{Z}}}$ and partition the matrix into $O(\log N)$ levels. At each level off-diagonal blocks representing non-self interactions are represented as low-rank or butterfly formats with a prescribed compression tolerance. Next, $\bar{\bar{\mathbf{Z}}}^{-1}$ is computed also with compressed representations using deterministic or randomized operations \cite{liu2020butterfly}. Once computed, a function $\bar{\mathbf{y}}=\bar{\bar{\mathbf{Z}}}^{-1}\bar{\mathbf{x}}$ with arbitrary input vector $\bar{\mathbf{x}}$ can be supplied to ARPACK. Note that $\bar{\bar{\mathbf{Z}}}^{-1}$ is approximately computed and preconditioned Krylov solvers may be needed for this function. However, preconditioned iterative solvers can significantly slow down the ARPACK computation \ylrev{due to significantly more numbers of matrix-vector multiplication required by a sequence of Arnoldi factorization from ARPACK} and instead we compute $\bar{\bar{\mathbf{Z}}}^{-1}$ with small compression tolerance and directly pass it to ARPACK without iterative solvers. It's worth mentioning that to properly reorder the matrix $\bar{\bar{\mathbf{Z}}}$, we associate each unknown with a Cartesian coordinate and use e.g., KD tree clustering algorithms \ylrev{or cobblestone-like schemes \cite{Shaeffer2007blr}}. For each flat/curvilinear triangle, there are $(p+3)(p+1)$ associated GWP($p$) bases $\bm{f}_n(\bm{r})$. We use the edge centers as the coordinates for the first $3(p+1)$ bases and triangle center for the next $p(p+1)$ bases. For each basis function $\hat{\bm{z}}\times\bm{e}_n(\bm{r})$, we use the port center as the coordinate. \ylrev{This list of $N_J+N_M$ coordinates is then passed to the clustering algorithm to compute the matrix reordering.}     
	
	Note that in the context of 3D surface IE, HODLR displays asymptotically unfavorable computational complexity for HOM modeling, as it employs both weak admissibility and low-rank compression (formats not intended for high-frequency Green's functions). However, HODLR remains highly efficient for the first few HOMs of a cavity when considering moderate system sizes (e.g., $N<20,000$), and as such, we employ HODLR as the default direct solver in IERD. For HOMs of very high frequencies, we can switch to HODBF whose CPU and memory complexities are asymptotically superior. See Section \ref{sec:pillbox} for numerical results.

	\section{Augmented Bayesian Optimization for Searching Resonance Frequencies \label{sec:BO}}
	The section describes the proposed efficient frequency search algorithm \rev{(Algorithm \ref{algo:optimizer})} that combines the global convergent behavior of BO and local refinement behavior of the down-hill simplex algorithm. 
	
	Assume that within the frequency band of interest $[f_{\min},f_{\max}]$ there are a total of $N_{mod}$ modes, and the modes indexed by $m$ are sorted by their resonance frequency $f^m$ in ascending orders. For each mode $m$, there is at most one eigenpair $(e^m(f),\bar{{\mathbf{I}}})$ returned by ARPACK at any frequency $f$. If no such eigenpair for mode $m$ exists at $f$, we simply let $e^m(f)=1$. Such a function $e^m(f)$ has only one global yet very sharp minimum at $f^m$, particularly for resonance modes without efficient power damping (i.e., with high Q factor). Efficient optimization algorithms need to efficiently locate the resonance regions with $e^m(f)<1$ and shrink to the global minimum $f^m$ for all resonance mode $m$.

	Let \{($\rev{f_i^m}, e^m_i$)\} be the samples of $e^m(f)$. In BO \cite{williams2006gaussian,luo2022sparse}, we assume a GP model $y^m(f)$ for $e^m(f)$ such that $y^m(f)\sim GP(\mu,\Sigma)$ with mean $\mu=0$ and covariance 
	\begin{align}
	\Sigma(f,f')=\sigma\exp(\frac{(f-f')^2}{l})+d\delta(f,f')\label{eq:gp}
	\end{align}
	where $\sigma,l,d$ are hyperparameters. From the samples \{($\rev{f_i^m}, e^m_i$)\}, these hyperparameters can be learned by optimizing the log-likelihood function of the GP model, known as the process of fitting a GP model to data. Once fitted, the GP surrogate can be used to quickly predict the mean and variance (or standard deviation) of $e^m(f)$ at any unknown frequency $f$. The mean and variance information can be used to define an acquisition function such as the Expected Improvement (EI), which is used to propose new trial frequency $f^*$. IERD will then evaluate $f^*$ with ARPACK to append the new sample $(f^*,e^m(f^*))$ to \{($\rev{f_i^m}, e^m_i$)\}, such as the Fun\_Eval function in Algorithm \ref{algo:optimizer}. Note that everytime called, Fun\_Eval will add a new sample to all detected modes so far, depending on the eigen pairs returned by ARPACK. More specifically, the returned eigenvectors and those from existing modes can be grouped with the mode tracking technique \cite{chen2015characteristic} by computing their inner products. 
	Two eigenvectors $\bar{\mathbf{I}}$ and $\bar{\mathbf{I}}^m$ are the same eigenmode only when $\bar{\mathbf{I}}^*\cdot\bar{\mathbf{I}}^m>\epsilon_{dot}$ for a prescribed threshold $\epsilon_{dot}$ ($\epsilon_{dot}=0.7-0.9$). Moreover, the recorded eigenvectors for existing modes $\bar{\mathbf{I}}^m$ are replaced by $\bar{\mathbf{I}}$ when $\bar{\mathbf{I}}^*\cdot\bar{\mathbf{I}}^m>\epsilon_{dot}$ and the new eigenvalue $e<\min_i{e^m_{i}}$ (see Line \ref{algo:line:updateI} of Algorithm \ref{algo:optimizer}). 
	Note that for each trial frequency $f$, the Fun\_Eval function will generate a sample $(f_i^m,e_i^m)$ for all modes, but only those with $e_i^m$ returned by ARPACK (i.e., $e_i^m<1$) denote a detected resonance.   
	
	The above procedure is repeatedly executed until the total number of affordable frequency trials per mode has reached a prescribed number $n_{BO}$ (See Line \ref{algo:line:BO} of Algorithm \ref{algo:optimizer}). The advantage of BO for IERD is that BO requires typically a much smaller number of trial frequencies compared with naive frequency sweep algorithms. That said, the objective function $e^m(f)$ for each mode exhibits a sharp global minimum and the smooth and stationary kernel in (\ref{eq:gp}) is not suited to detect the true \rev{non-smooth} minimum \rev{\cite{luo2021non,luo2021hybrid}}.  
	
\begin{figure}[!t]
		\centering
		\includegraphics[width=0.6\columnwidth]{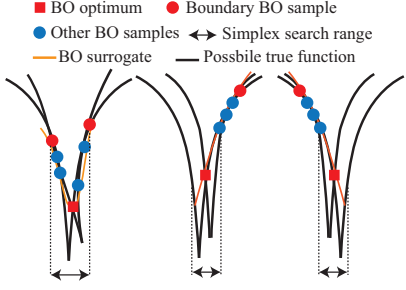}
		\caption{Identification of the search range for down-hill simplex based on the BO samples $\{(f_{i}^m,e^m_{i})\}$, particularly, the three possible relative locations of optimal ($f_{\min}^m=\argmin_ie^m_i$) and boundary BO samples ($\min{f_{i}^m}$ and $\max{f_{i}^m}$) shown in the left, middle and right.  \label{fig:simplexrange}}
	\end{figure}

	To improve the search quality, we augment BO with a down-hill simplex refinement inside each resonance region \rev{using BO samples ${(f_{i}^m,e^m_{i})}$} (see Line \ref{algo:line:SX} of Algorithm \ref{algo:optimizer}). The down-hill simplex optimization is a very efficient derivative-free local algorithm requiring a small number of function samples in our application. However, this algorithm heavily rely on good initial guesses \rev{and search regions}. Given that the objective functions $e^m(f)$ has a pre-defined ``V''-like shapes, we can leverage BO samples to define the lower and upper frequency bounds for the simplex search. As an illustration, Fig. \ref{fig:simplexrange} shows three possible configurations of the boundary (shown in red circles) and optimal (shown in red squares) BO samples and potential locations of the global minimum. Note that standard GP surrogate model in BO (shown in orange curves) cannot represent the true objective function sufficient accuracy near a non-smooth optimum. \rev{More specifically, from the BO samples ${(f_{i}^m,e^m_{i})}$ of one mode let $f_{\min}^m=\argmin_ie^m_i$. If $f_{\min}^m$ appears in between $\min{f_{i}^m}$ and $\max{f_{i}^m}$, the simplex search range is $(f_{\min}^{SX},f_{\max}^{SX})=(\min{f_{i}^m},\max{f_{i}^m})$; otherwise if $f_{\min}^m$ equals $\min{f_{i}^m}$, the simplex search range is 
		$(f_{\min}^{SX},f_{\max}^{SX})=((1-\alpha)f_{\min}^m,f_{n_r})$ where $\alpha=0.004$ is a small constant and $f_{n_r}$ is the first trial frequency to the right of $f_{\min}^m$; the case that $f_{\min}^m$ equals $\max{f_{i}^m}$ is treated in a similar fashion. Together with these search ranges, we can use their midpoints  $(f_{\min}^{SX}+f_{\max}^{SX})/2$ as the initial guess to down-hill simplex.} See Line \ref{algo:line:SX} of Algorithm \ref{algo:optimizer} for more details.

	\begin{algorithm}[t!]
		\small
		\caption{Hybrid BO and down-hill simplex optimization for frequency search}
		\label{algo:optimizer}
		\begin{algorithmic}[1]
			\State \textbf{Input:} 
			\State $n_{BO}$: number of trial frequencies per mode in BO
			\State $n_{SX}$: number of trial frequencies per mode in down-hill simplex
			\State $n_{AK}$: number of smallest eigenvalues for APRACK
			\State $(f_{\min},f_{\max})$: lower and upper bounds for the search
			\State \textbf{Output:}
			\State $N_{mod}$: number of resonance modes (initialized to 0)
			\State ${(f_{i}^m,e^m_{i})}$: frequencies and the eigenvalues of each mode $m\leq N_{mod}$ for all trials $i$. $f^m=\argmin_ie_{i}^m$ is the resonance frequency of mode $m$.
			
			\Statex
			\State \textbf{BO:}  \label{algo:line:BO}
			\State Generate $n_{BO}'=n_{BO}/2$ random frequencies in [$f_{\min},f_{\max}$]. For each frequency sample $f$, call Fun\_Eval($f$,$N_{mod}$)
			\State \Comment{\rev{The number of initial random samples $n_{BO}'=n_{BO}/2$ is a heuristic choice.}}
			\For{$n=1$ to $n_{BO}-n_{BO}'$}		
			\For{each existing mode $m\leq N_{mod}$}
			\State Fit a GP model using data $\{(f_{i}^m,e^m_{i})\}$  
			\State Use the EI function to propose the next trial frequency $f$
			\State Fun\_Eval($f$,$N_{mod}$) 
			\EndFor
			\EndFor
			
			\Statex
			\State \textbf{Downhill-simplex:} \label{algo:line:SX}
			\For{each existing mode $m\leq N_{mod}$}
			\State $f_{\min}^m=\argmin_ie^m_i$ 
			\If{$\min{f_{i}^m}<f_{\min}^m<\max{f_{i}^m}$}
			\State $(f_{\min}^{SX},f_{\max}^{SX})=(\min{f_{i}^m},\max{f_{i}^m})$
			\ElsIf{$\min{f_{i}^m}==f_{\min}^m$}
			\State $(f_{\min}^{SX},f_{\max}^{SX})=((1-\alpha)f_{\min}^m,f_{n_r})$ \Comment{$f_{n_r}$ denotes the right neighbor of $f_{\min}^m$, and $\alpha=0.004$ is a small constant.}
			\ElsIf{$\max{f_{i}^m}==f_{\min}^m$}
			\State $(f_{\min}^{SX},f_{\max}^{SX})=(f_{n_l},(1+\alpha)f_{\min}^m)$ \Comment{$f_{n_l}$ denotes the left neighbor of $f_{\min}^m$}
			\EndIf
			\State Call down-hill simplex algorithm with search range $[f_{\min}^{SX},f_{\max}^{SX}]$ for $n_{SX}$ samples. Each sample calls Fun\_Eval($f$,$N_{mod}$).  
			\EndFor

			\Statex
			\Procedure{Fun\_Eval}{$f$,$N_{mod}$} \label{algo:line:func_eval}
			\State Call ARPACK with HODLR/HODBF for $n_{AK}$ smallest eigenvalues. Ignore eigen pairs $(e,\bar{{\mathbf{I}}})$ with  $||\bar{\mathbf{I}}||_1/||\bar{\mathbf{I}}||_\infty<\epsilon_{norm}$. 
			\For{each existing mode $m\leq N_{mod}$ with eigen vector $\bar{{\mathbf{I}}}^m$}
			\If{$\exists$ eigen pair $(e,\bar{{\mathbf{I}}})$ from ARPACK such that $\bar{\mathbf{I}}^*\cdot\bar{\mathbf{I}}^m>\epsilon_{dot}$}
			\State Let $\bar{{\mathbf{I}}}^m=\bar{{\mathbf{I}}}$ if $e<\min_i{e^m_{i}}$\label{algo:line:updateI}
			\State Remove $(e,\bar{{\mathbf{I}}})$ from ARPACK results and Append $(f,e)$ to $\{(f_{i}^m,e^m_{i})\}$    
			\Else
			\State Append $(f,1)$ to $\{(f_{i}^m,e^m_{i})\}$     
			\EndIf		
			\EndFor	
			
			\For{each remaining eigen pair $(e,\bar{{\mathbf{I}}})$ from ARPACK}	
			\State $N_{mod}=N_{mod}+1$
			\State Let $(f_1^{N_{mod}},e_1^{N_{mod}})=(f,e)$ and $\bar{{\mathbf{I}}}^{N_{mod}}=\bar{{\mathbf{I}}}$
			\EndFor

			\EndProcedure
		\end{algorithmic}
	\end{algorithm}

	%
	%
	%
	%

	In our implementation, we use two Python packages: the GPTune software \cite{gptune} for the BO optimization and the SciPy software \cite{2020SciPy-NMeth} for the down-hill simplex algorithm. For each trial frequency, the IE-based linear eigensolver (i.e., ButterflyPACK and ARPACK) is invoked from the Python code with multiple MPIs ranks. Overall, the runtime for the proposed IERD framework is at most $(n_{BO}+n_{SX})N_{mod}T_{eigen}$, where $T_{eigen}$ is the CPU time for each trial frequency $f$.   	
	
	\section{Numerical Results}\label{sec:result}
	This section provides several numerical examples to demonstrate the efficiency, accuracy and robustness of the proposed IERD framework for modeling resonance modes in both canonical and realistic RF accelerator cavities. When the cavity involves waveguide ports with irregular-shaped cross sections, we use CST Studio Suite \cite{solvers2020cst}'s 2D eigenmode solver to precompute the basis functions $\bm{z}\times \bm{e}_j(\bm{r})$. As the reference, we also show computational results using CST's 3D eigenmode solver \cite{solvers2020cst} and the SLAC Omega3P eigenmode solver \cite{lee2009omega3p,xiao2019advances}, both of which are state-of-the-art FEM solvers for electromagnetic simulations. For Omega3P, we use the quadratic eigenproblem formulation as described in \cite{lee2009omega3p}. 
	
	All numerical experiments are performed on the Cori Haswell system at NERSC. Cori Haswell is a Cray XC40 system and consists of 2388 dual-socket nodes with Intel Xeon E5-2698v3 processors running 16 cores per socket. The nodes are equipped with 128 GB of DDR4 memory and are connected through the Cray Aries interconnect. 
	\subsection{Pillbox cavity with no port}\label{sec:pillbox}
\begin{figure}[!t]
		\centering
		\includegraphics[width=0.8\columnwidth]{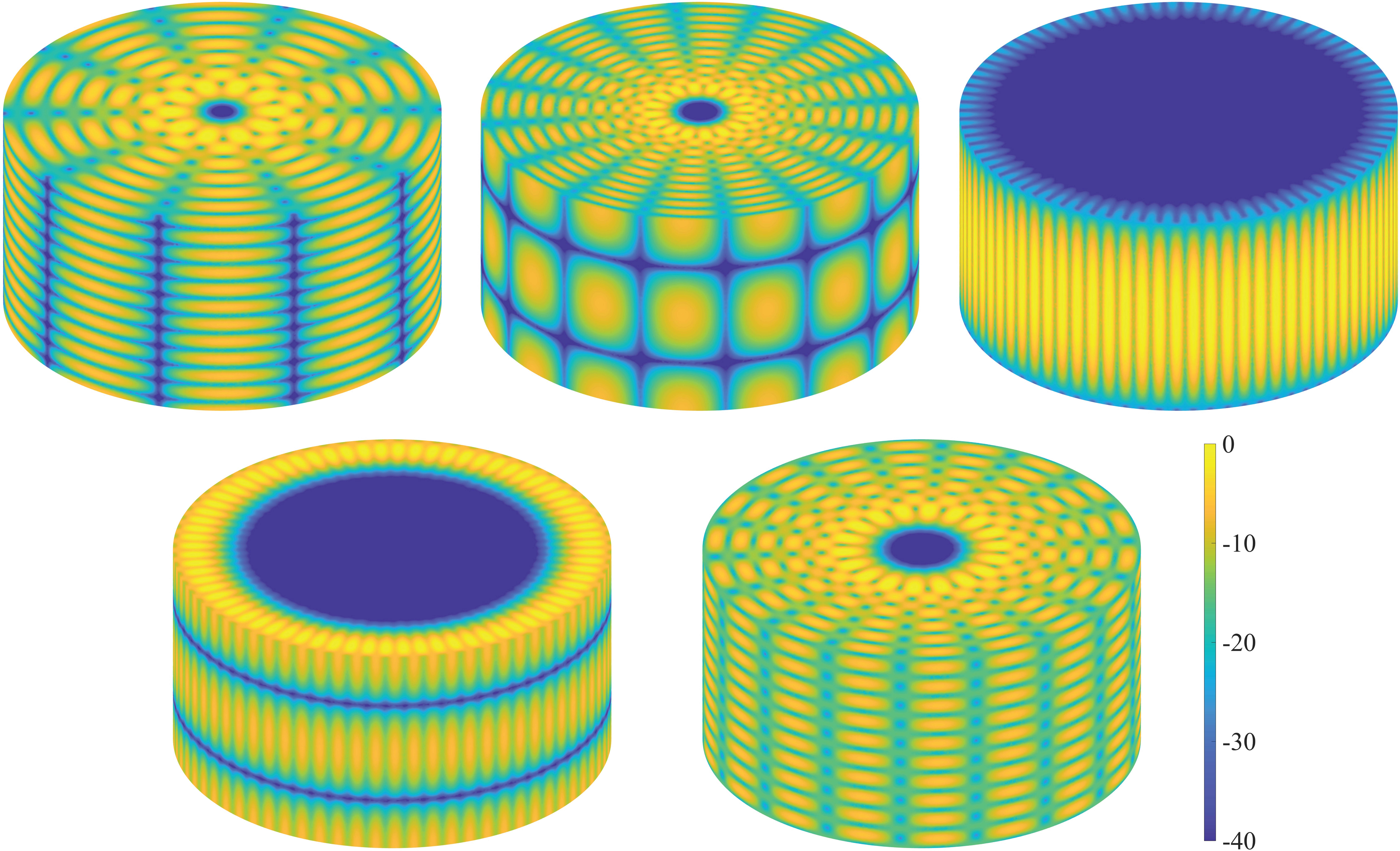}
		\caption{	
			$|\bm{J}|$ (in dB) for the five resonance modes of the pillbox cavity model detected at $f$=20.472 GHz with the proposed solver.\label{fig:pillboxmodes}}
	\end{figure}

	In this subsection, we use a canonical pillbox cavity of radius 0.1 m and height 0.1 m to demonstrate the computational efficiency of the proposed IERD framework. First, we validate the accuracy of IERD by setting $f=$20.472 GHz, one of the HOM frequencies without frequency search. This leads to a diameter of $13.6~\lambda$ and we use a mesh of $213,792$ triangle elements and $N=N_J=320,688$ RWG basis functions. When no port is considered, the system matrix $\bar{\bar{\mathbf{Z}}}$ reduces to $\bar{\bar{\mathbf{T}}}$ of (\ref{eq:IEmatrix}). The inverse $\bar{\bar{\mathbf{Z}}}^{-1}$ is computed with the HODLR direct solver with compression tolerance $\epsilon=10^{-4}$ with 16 Cori nodes and supplied to ARPACK for the smallest 10 eigenvalues. The five detected eigenmodes $\bm{J}$ corresponding to valid resonance are plotted in Fig. \ref{fig:pillboxmodes}, which agree with the analytical solutions. 
	
	Next, different solver options are compared to demonstrate the efficiency of IERD. In addition to the HODLR direct solver with RWG basis functions, we also tested the HODBF direct solver with RWG basis functions, and the HODLR direct solver with GWP(1) basis functions and curvilinear elements. The system size $N$ is kept similar across the three solvers and technical data is listed in Table \ref{tab:tech_data}. Note that the use of GWP(1) basis on a coarser mesh can lead to significant reduction in CPU time. \ylrev{This is mainly due to that the time for compressing $\bar{\bar{\mathbf{Z}}}$ is dominated by that of computing entries of submatrices of $\bar{\bar{\mathbf{Z}}}$ required by the adaptive cross approximation (ACA)\cite{liu2019parallelACA}. Given similar numbers of unknowns $N$ (and similar ranks), GWP(1) requires a much less number of triangles than RWG. As the matrix is assembled on a triangle basis, the entry evaluation using GWP(1) becomes much faster.} Also, the use of HODBF instead of HODLR can significantly reduce the memory usage for very high-order HOM modeling \ylrev{as HODBF can achieve significantly better compression performance than HODLR for high-frequency problems \cite{sayed2022VIE}}. As the reference, we also used the Omega3P solver on 16 Cori nodes to detect resonance modes around $20.5$ GHz with a tetrahedron mesh of similar mesh density. This leads to a FEM matrix of 17 million unknowns and the CPU time is similar to our IERD framework. That said, IERD with HODBF requires only 51.1 GB memory while Omega3P requires 1.56 TB memory.   
	
	\begin{table}[tp!]
		\footnotesize
		\caption{Comparison of the HODLR and HODBF-enhanced eigen solver for the pillbox cavity model with $f$=20.472 GHz using 16 Cori Haswell nodes. }
		\begin{center}
			\begin{tabular}{|c|c|c|c|c|}
				\hline			
				Algorithm & HODLR & HODBF  & HODLR & Omega3P \\
				\hline
				Diameter & 13.6 $\lambda$ & 13.6 $\lambda$ & 13.6 $\lambda$ & 13.6 $\lambda$  \\
				\hline				
				\#of triangles & 213,792 & 213,792 & 53,240 & -  \\										
				\hline
				\#of tetrahedrons & - & - & - & 14,861,793  \\										
				\hline	
				Max edge length & 2.1 mm& 2.1 mm & 4 mm & 2.7 mm  \\			
				\hline									
				Basis & RWG & RWG & GWP($1$) & Nedelec\\				
				\hline				
				\#DoFs & 320,688 & 320,688 & 266,200 & 17,185,459 \\							
				\hline				
				Compression tolerance & $10^{-4}$ & $10^{-4}$ & $10^{-4}$ & - \\				
				\hline				
				Assembly\ylrev{\&compress} & 39 min &  8 min & 12 min &  16.8 min \\				
				\hline						
				Max. rank & 4113 &  380 & 4037 &  - \\		
				\hline								
				Inversion time & 54 s & 38 min & 52 s & -  \\			
				\hline
				Total memory  & 250 GB & 51.1 GB & 200.8 GB & 1.56 TB  \\
				\hline				
				Eigen solver time & 13.4 s & 49.2 s  & 11.9 s & 7.4 min   \\
				\hline
				\#of detected modes & 5 & 5  & 5 & 5   \\
				\hline									
			\end{tabular}
		\end{center}
		\label{tab:tech_data}
	\end{table}

	\begin{figure}[!t]
		\centering
		\includegraphics[width=\columnwidth]{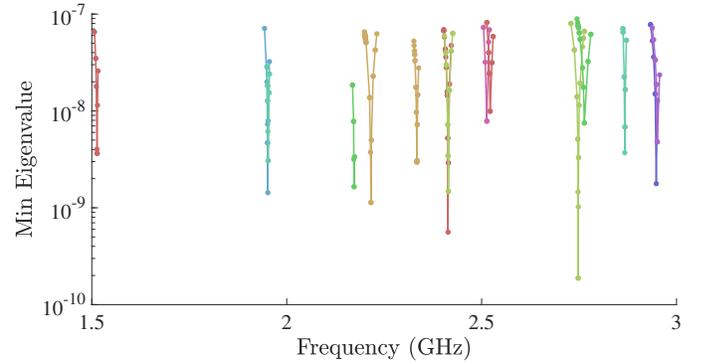}
		\caption{	
			Eigenvalues of the first 15 resonance modes in [1.5 GHz, 3 GHz] for the single-cell cavity computed by the proposed solver on 16 Cori Haswell nodes. The rectangular waveguides are treated as closed boundaries. Each color represents one detected resonance with trial frequencies (the dots) suggested by the proposed solver. The detected resonance frequency for each mode is the frequency where the eigenvalue attains the minimum. \rev{Note that the frequency samples $f_i^m$ with $e_i^m=1$ are not plotted.}\label{fig:cavity_rec_fullspectrum}}
	\end{figure}
	
	\begin{figure}[!t]
		\centering
		\includegraphics[width=\columnwidth]{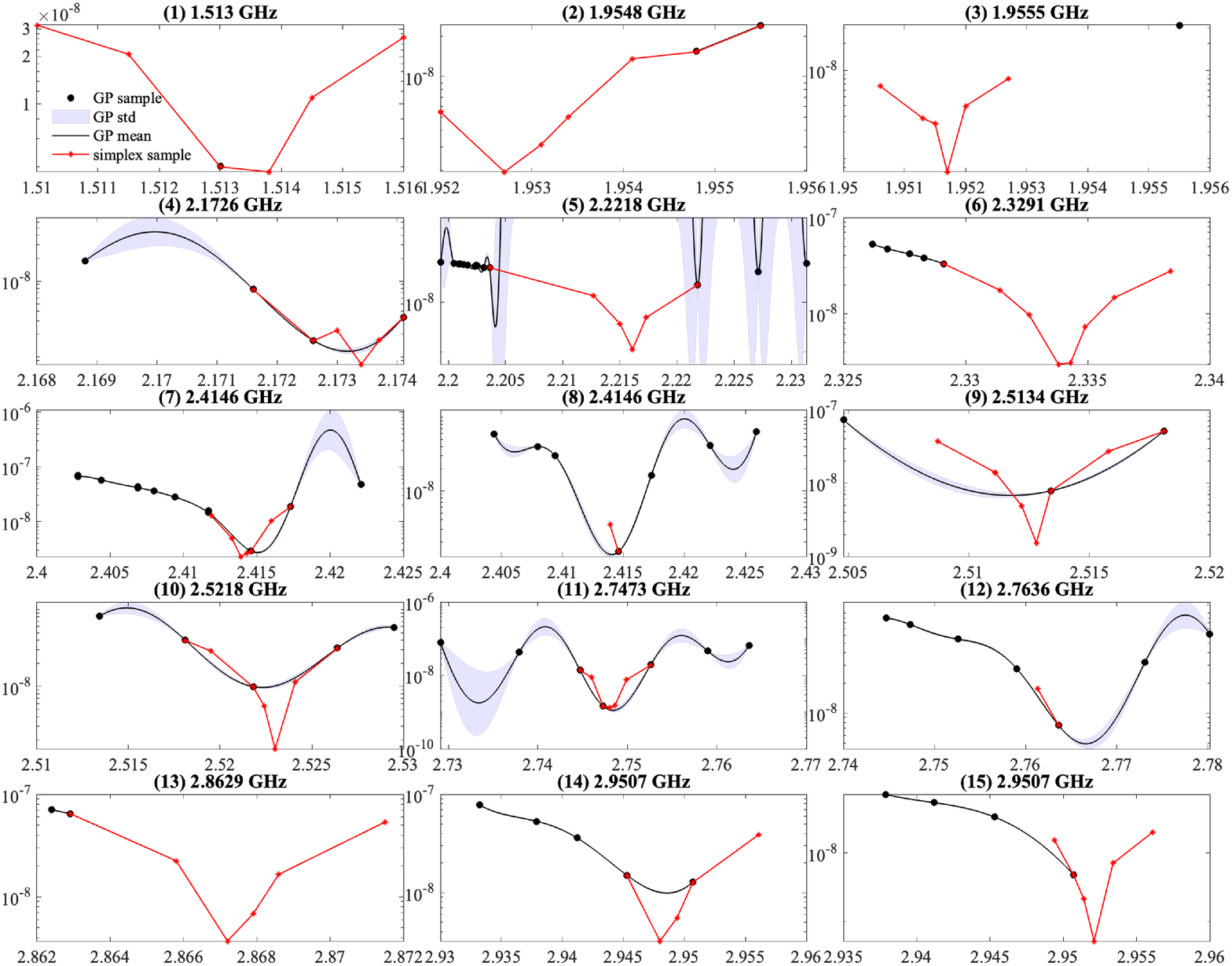}
		\caption{	
			Zoomed view of the 15 resonance modes in Fig. \ref{fig:cavity_rec_fullspectrum}. Each subfigure shows one resonance mode numbered by its resonance frequency. \rev{The black dots, curves and blue shaded areas represent the samples, predicted mean and predicted standard deviation (std) by the Bayesian optimization, respectively. The red curves represent samples generated by the downhill simplex optimization.} Note that the sensitivity or Q factor of the resonance is reflected by the bandwidth of the data.\label{fig:cavity_rec_zoomspectrum}}
	\end{figure}
	
	\begin{figure}[!t]
		\centering
		\includegraphics[width=\columnwidth]{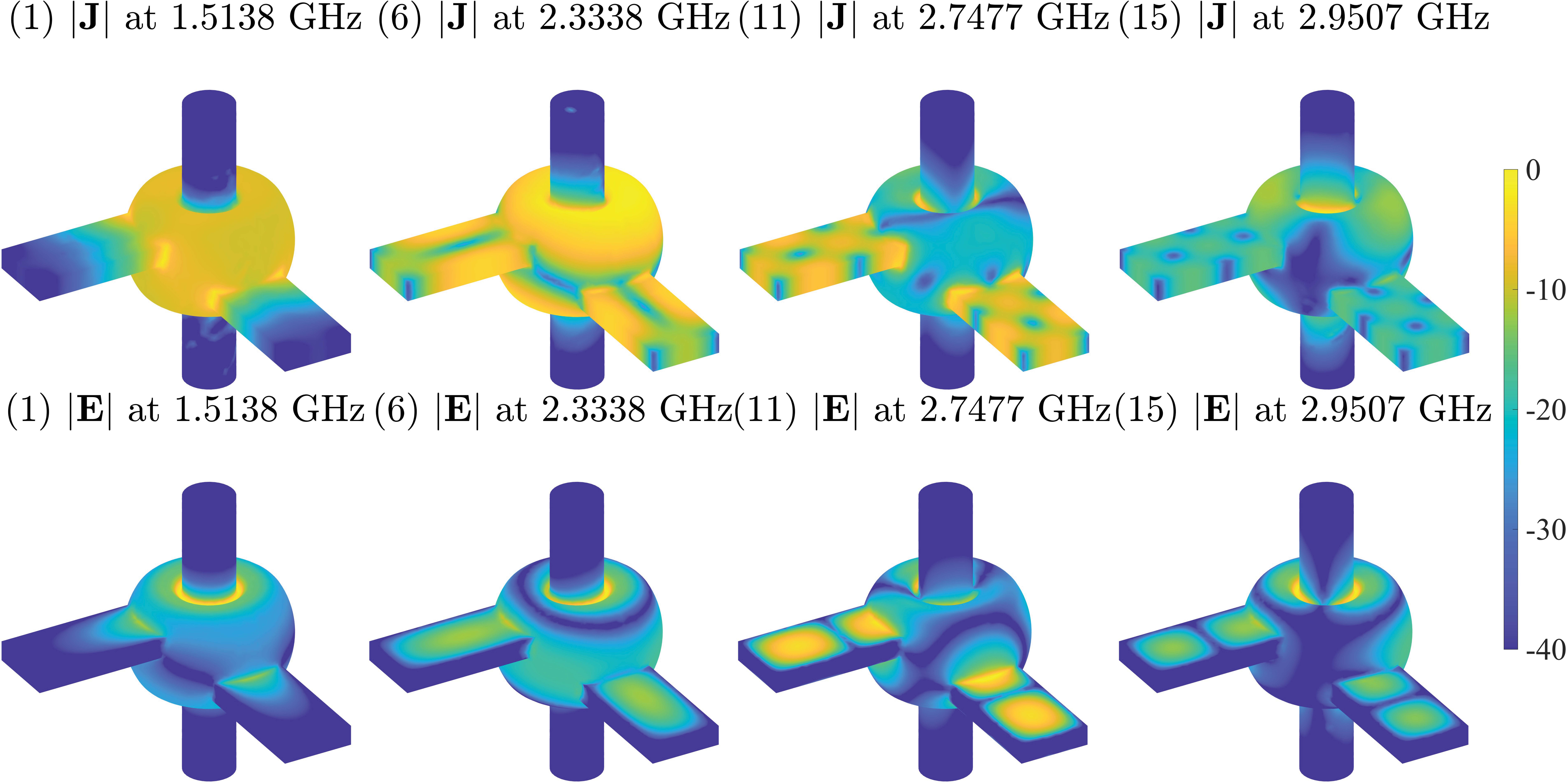}
		\caption{	
			$|\bm{J}|$ and surface $|\bm{E}|$ (in dB) for the $1^{st}$, $6^{th}$, $11^{th}$, and $15^{th}$ resonance modes for the single-cell cavity with closed boundaries at the ports computed by the proposed solver. Note that the $1^{st}$ mode represents the working mode. \label{fig:EJ_cavity_rec}}
	\end{figure}

	\begin{figure}[!t]
		\centering
		\includegraphics[width=\columnwidth]{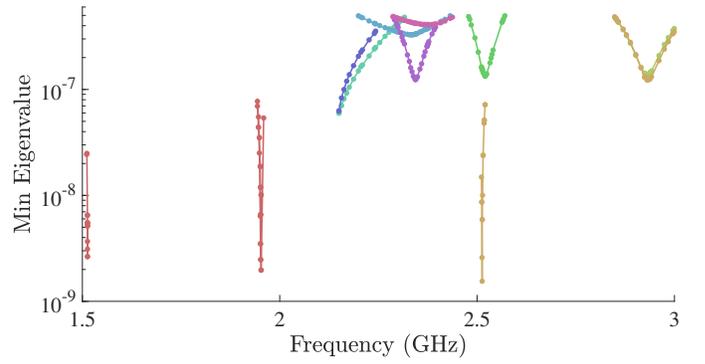}
		\caption{	
			Eigenvalues of the first 12 resonance modes in [1.5 GHz, 3 GHz] for the single-cell cavity with two rectangular waveguide ports computed by the proposed solver on 16 Cori Haswell nodes. Each color represents one detected resonance with trial frequencies (the dots) suggested by the proposed solver. The detected resonance frequency for each mode is the frequency where the eigenvalue attains the minimum. Note that most HOMs exhibit larger eigenvalues and higher bandwidth as the power is efficiently damped by the ports. \label{fig:cavity_rec_port_fullspectrum}}
	\end{figure}
	
	\begin{figure}[!t]
		\centering
		\includegraphics[width=\columnwidth]{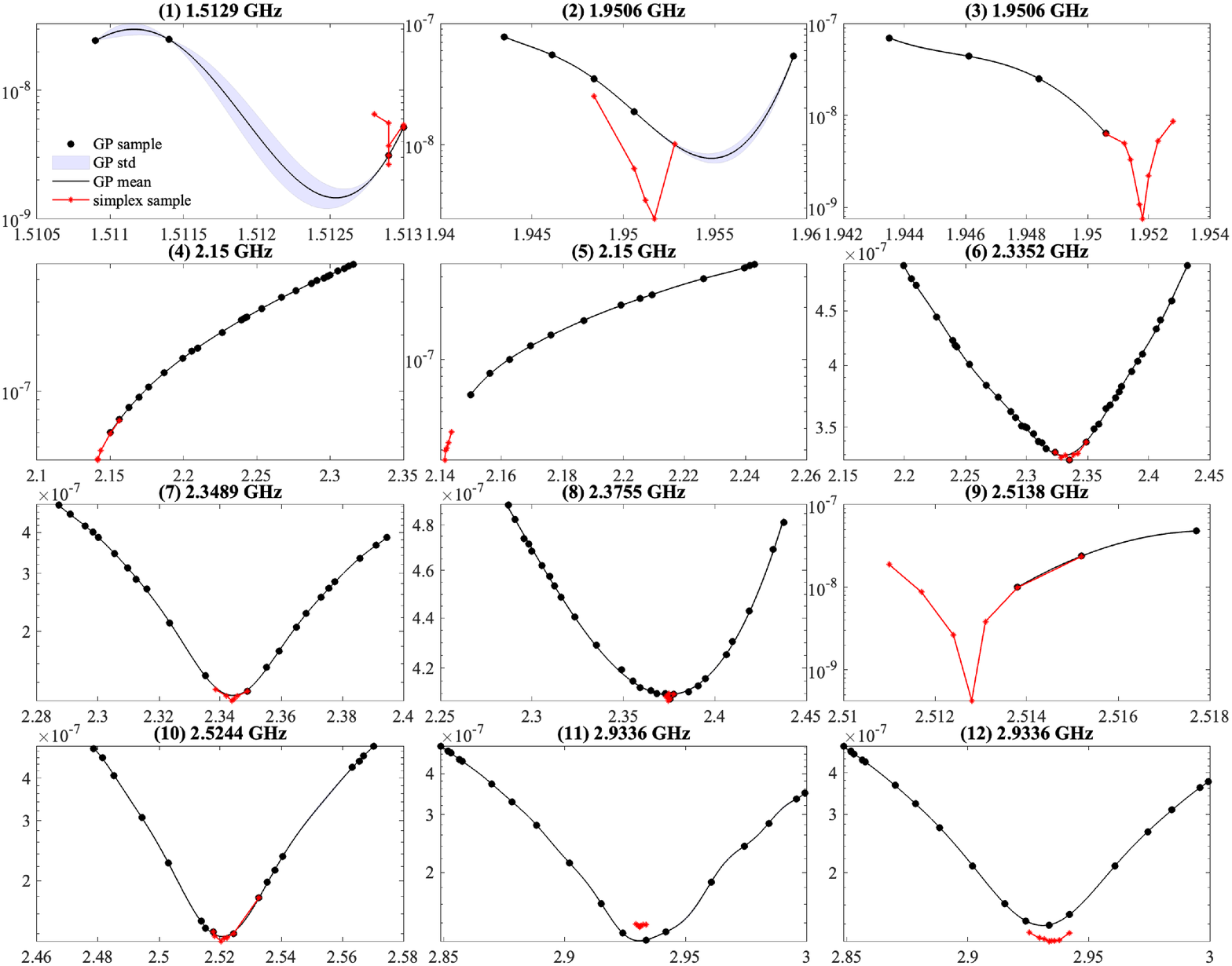}
		\caption{	
			Zoomed view of the 12 resonance modes in Fig. \ref{fig:cavity_rec_port_fullspectrum}. Each subfigure shows one resonance mode numbered by its resonance frequency. \rev{The black dots, curves and blue shaded areas represent the samples, predicted mean and predicted standard deviation (std) by the Bayesian optimization, respectively. The red curves represent samples generated by the downhill simplex optimization.} Note that the sensitivity or Q factor of the resonance is reflected by the bandwidth of the data.\label{fig:cavity_rec_port_zoomspectrum}}
	\end{figure}
	
	\begin{figure}[!t]
		\centering
		\includegraphics[width=\columnwidth]{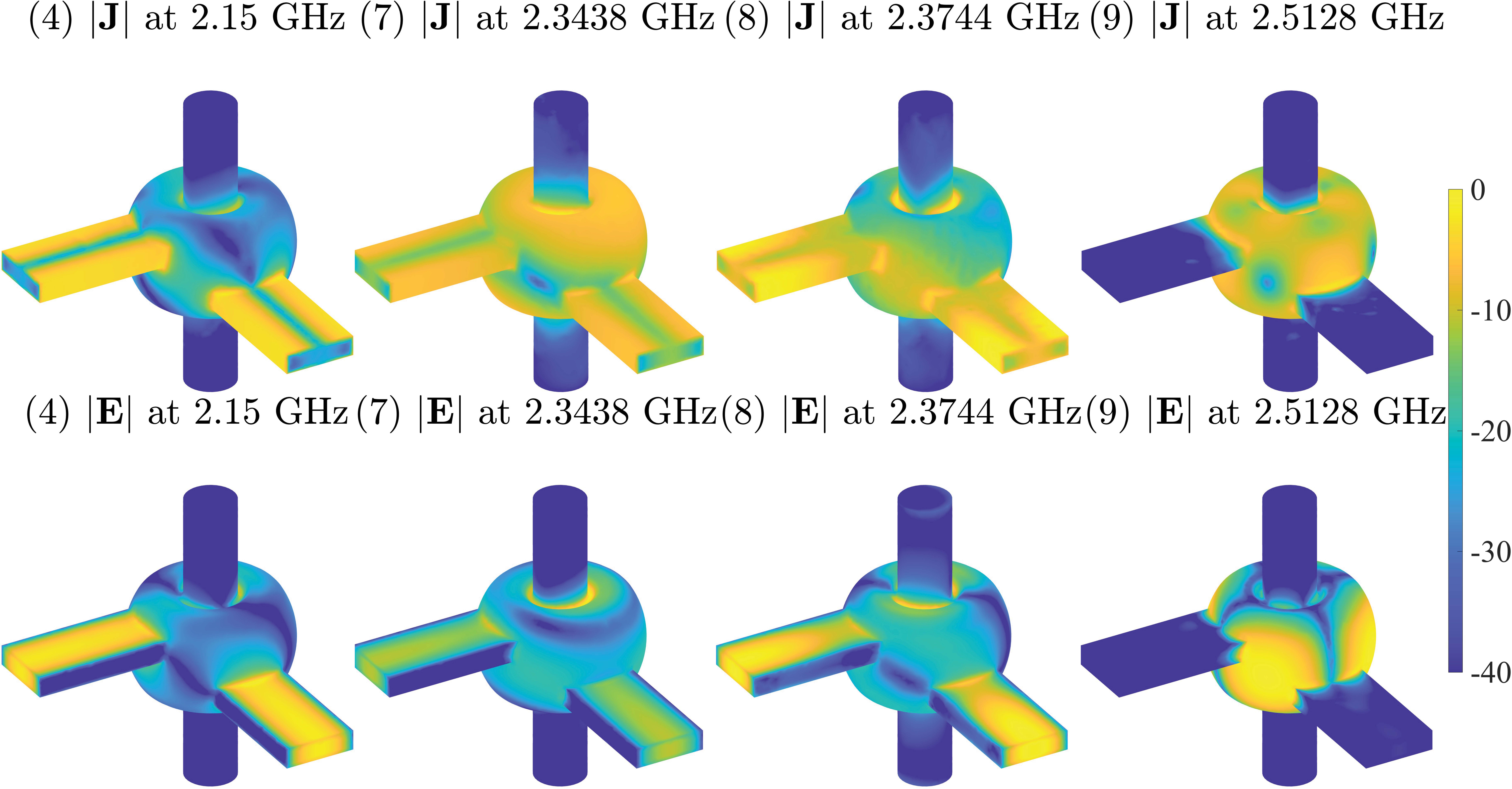}
		\caption{	
			$|\bm{J}|$ and surface $|\bm{E}|$ (in dB) for the $4^{th}$, $7^{th}$, $8^{th}$, and $9^{th}$ HOMs for the single-cell cavity with open boundaries at the ports computed by the proposed solver. Note that the $4^{th}$, $7^{th}$, and $8^{th}$ modes show efficient power damping. \label{fig:EJ_cavity_rec_port}}
	\end{figure}

	\begin{figure}[!t]
		\centering
		\includegraphics[width=\columnwidth]{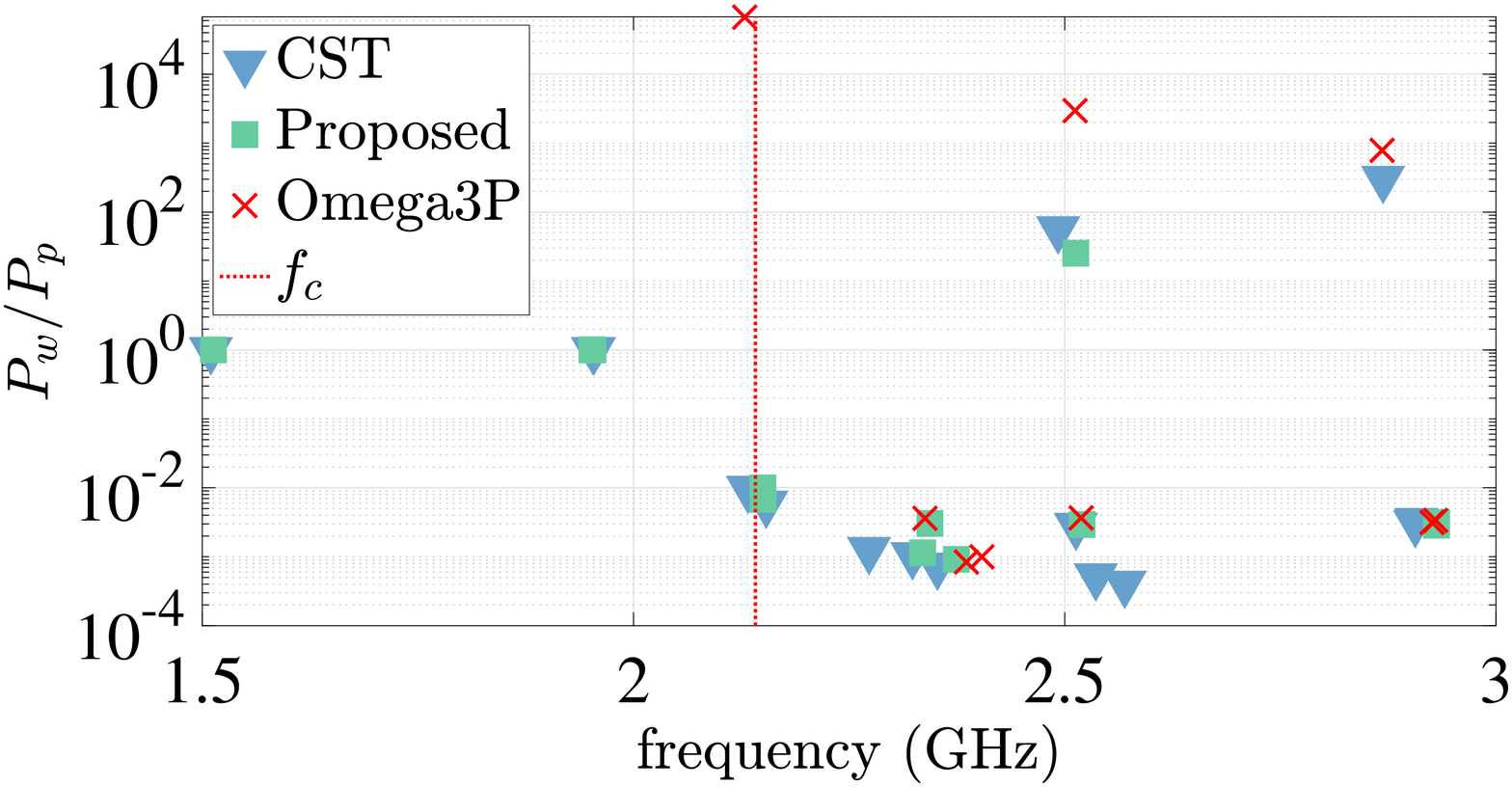}
		\caption{	
			The power ratios $P_w/P_p$ of all resonance modes in [1.5 GHz, 3 GHz] detected by CST and the proposed solver, and for all resonance modes in [2.141 GHz, 3 GHz] detected by Omega3P, for the single-cell cavity with two rectangular waveguide ports. The red vertical line denotes the lowest cutoff frequency $f_c=2.141$ GHz of the ports, below which $P_p=0$ and we simply set $P_w/P_p=1$. \label{fig:cavity_rec_ratio}}
	\end{figure}

	\subsection{Single-cell cavity with two rectangular waveguide ports}
	In this subsection, we demonstrate the efficacy of IERD for a realistic single-cell cavity with two rectangular waveguide ports, a simulation model for the third harmonic cavity in the Advance Light Source at Lawrence Berkeley National Laboratory \cite{als3hc}. The two ports have an identical cross section and mode vectors for the waveguide are $\bm{e}_j=\frac{m}{b}\cos\frac{n\pi x}{a}\sin\frac{m\pi y}{b}\hat{\bm{x}}-\frac{n}{a}\sin\frac{n\pi x}{a}\cos\frac{m\pi y}{b}\hat{\bm{y}}$ for $\mathrm{TE}_{nm}$ modes and $\bm{e}_j=\frac{n}{a}\cos\frac{n\pi x}{a}\sin\frac{m\pi y}{b}\hat{\bm{x}}+\frac{m}{b}\sin\frac{n\pi x}{a}\cos\frac{m\pi y}{b}\hat{\bm{y}}$ for $\mathrm{TM}_{nm}$ modes, where $x,y$ and $\hat{\bm{x}},\hat{\bm{y}}$ are local coordinates, and $a=70$ mm, $b=20$ mm in this model. The lowest cutoff frequency is $f_c=2.141$ GHz of the $\mathrm{TE}_{10}$ mode, and we search the resonance modes in [1.5 GHz, 3 GHz]. Therefore $N_M=0$ when $f<f_c$, and $N_M=2$ when $f\geq f_c$ (i.e., each port supports at most one propagating mode). The model discretization leads to $N_J=25,455$ RWG basis functions and ARPACK computes the smallest 20 eigenvalues. Each trial frequency $f$ requires about 35 s CPU time \ylrev{when $n_{AK}$=20} and 5 GB memory using HODLR on 16 Cori nodes.  
	First, we test the proposed frequency searching algorithm by setting $N_M=0$ regardless of the trial frequency (i.e., the waveguide ports are treated as closed boundaries). Fig. \ref{fig:cavity_rec_fullspectrum} plots the eigenvalues corresponding to the first 15 modes each represented by one color. Since closed boundaries are used, each mode exhibits a narrow bandwidth and a small minimum eigenvalue. Fig. \ref{fig:cavity_rec_zoomspectrum} plots the eigenvalues for each mode separately. The black dots and red dots represent trial frequencies selected by the Bayesian optimization and down-hill simplex algorithm, respectively. Clearly, Bayesian optimization can locate the resonance region for all modes and down-hill simplex can effectively refine the resonance frequencies to find the sharp minimum. The detected resonance frequencies are labeled in the title of each subfigure. Fig. \ref{fig:EJ_cavity_rec} shows the $|\bm{J}|$ and surface $|\bm{E}|$ for the $1^{st}$, $6^{th}$, $11^{th}$, and $15^{th}$ modes where $1^{st}$ mode is the working mode. Here $\bm{E}$ is simply computed by $\bm{E}=\frac{\nabla\cdot\bm{J}}{\imath\omega\epsilon_0}$. Note that $|\bm{E}|=0$ on the ports. 
	
	Next, we test the proposed frequency search algorithm with $N_M=2$ for $f\geq f_c$. Fig. \ref{fig:cavity_rec_port_fullspectrum} plots the eigenvalues corresponding to the first 12 modes each represented by one color. Note that the first three modes are below $f_c$ and $9^{th}$ mode is not damped efficiently via the wavguide ports. Therefore they exhibit narrow bandwidths and small minimum eigenvalues. All the other eight modes are damped efficiently and they exhibit higher bandwidths and larger minimum eigenvalues. Fig. \ref{fig:cavity_rec_port_zoomspectrum} plots the eigenvalues for each mode separately. The black dots and red dots represent trial frequencies selected by the Bayesian optimization and down-hill simplex algorithm, respectively. For the aforementioned four modes (subfigures (1)(2)(3)(9)) simplex refinement is critical for identifying more accurate resonance frequencies. That said, for the other modes whose power is damped efficiently, Bayesian optimization itself can yield sufficiently accurate solutions. Fig. \ref{fig:EJ_cavity_rec_port} shows the $|\bm{J}|$ and surface $|\bm{E}|$ for the $4^{th}$, $7^{th}$, $8^{th}$, and $9^{th}$ modes. Note that the $9^{th}$ mode doesn't have sufficient power damping. 
	
	Furthermore, we measure the damping efficiency of each mode by computing the ratio $P_w/P_p$, where $P_w=\frac{1}{2}\mathrm{Re}\{<\bm{J}^*,Z_s\bm{J}>\}$ is the power dissipated at the cavity wall, and $P_p=\frac{1}{2}\mathrm{Re}\{<\bm{J}^*,\bm{M}>\}$ is the power damped from the waveguide ports. The ratio is plotted in Fig. \ref{fig:cavity_rec_ratio} for the 12 resonance modes. For modes detected below $f_c$ we simply set $P_w/P_p=1$ for plotting purpose. As the reference, the ratios for the modes detected by CST and Omega3P are also plotted. It's worth mentioning that in this comparison study, for modes above $f_c$, CST doesn't yield accurate resonance frequencies. In contrast, Omega3P results agree better with IERD for $f>f_c$. That said, Omega3P cannot compute resonance modes accurately for $f\leq f_c$ and $f\approx f_c$ when waveguide port boundaries are used. For example, see the ratio for the mode near $f=f_c$.

	\subsection{Five-cell cavity with six H-shaped  waveguide ports}
	In this subsection, the applicability of IERD to complicated cavity geometry with irregular-shaped ports are demonstrated using a five-cell cavity model with six H-shaped waveguide ports with identical cross sections, a simulation model for the Energy Recovery LINAC in Electron Ion Collider at Brookhaven National Laboratory ~\cite{ERL-EIC}.

\begin{figure}[!t]
		\centering
		\includegraphics[width=0.8\columnwidth]{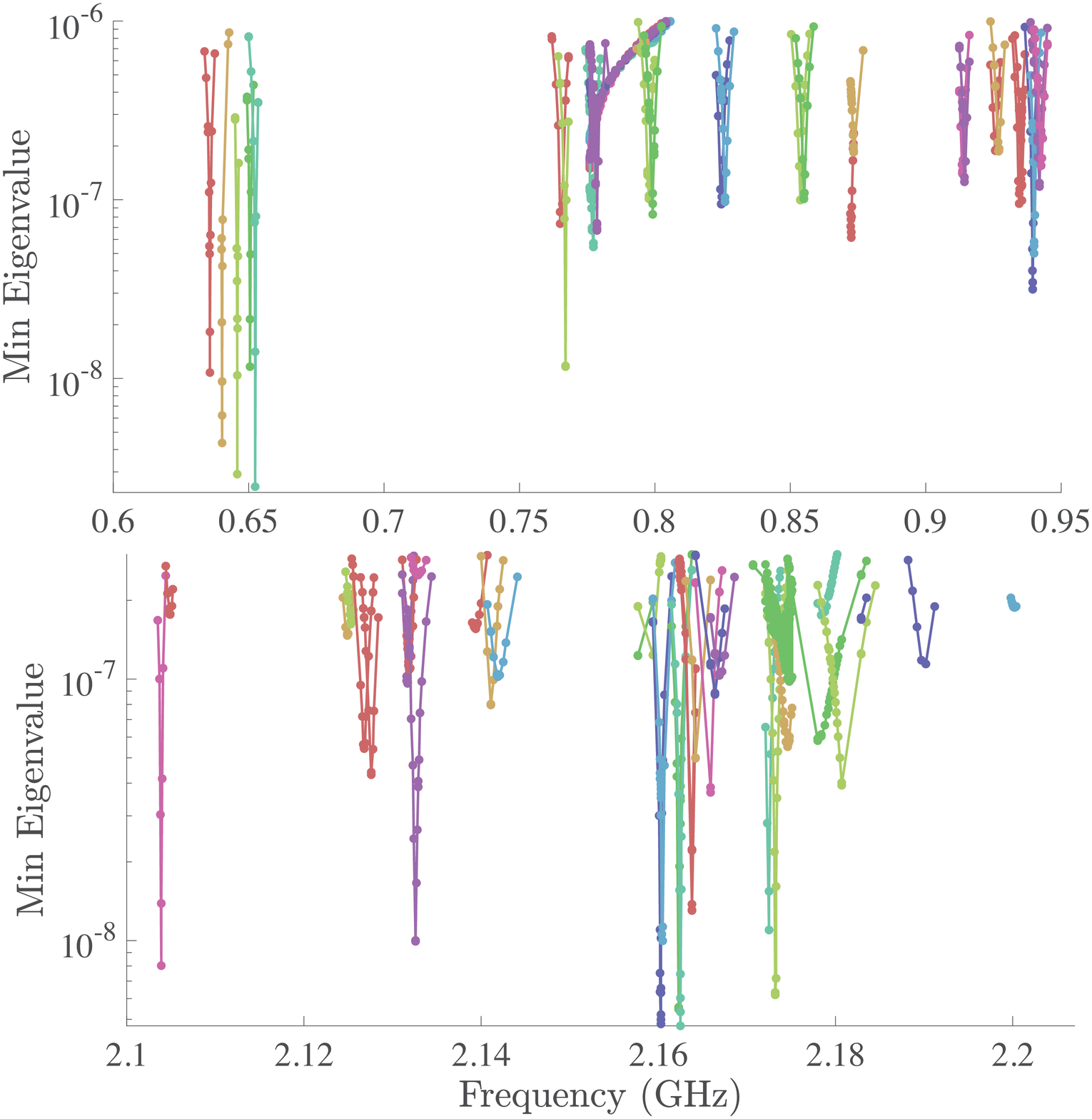}
		\caption{	
			Eigenvalues of the first 22 resonance modes in [0.6 GHz, 1 GHz] and the first 35 resonance modes in [2.1 GHz, 2.21 GHz] for the 5-cell cavity with 6 H-shaped waveguide ports computed by the proposed solver on 32 Cori Haswell nodes. Each color represents one detected resonance with trial frequencies (the dots) suggested by the proposed solver. The detected resonance frequency for each mode is the frequency where the eigenvalue attains the minimum. Note that most HOMs exhibit larger eigenvalues and higher bandwidth as power is efficiently damped by the ports.  \label{fig:cavity_5cell_port_fullspectrum}}
	\end{figure}

\begin{figure}[!t]
		\centering
		\includegraphics[width=\columnwidth]{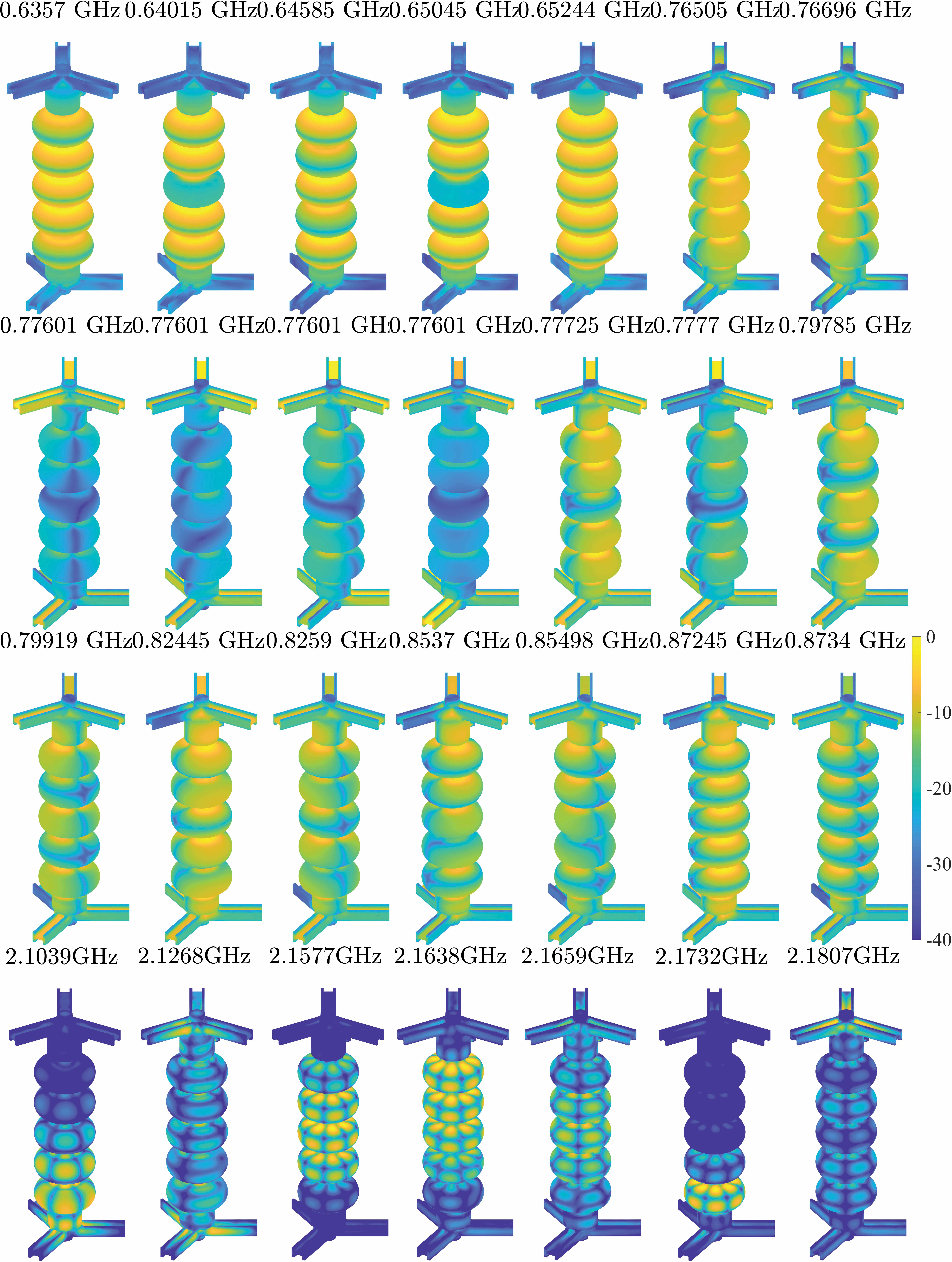}
		\caption{	
			surface $|\bm{E}|$ (in dB) for the first 14 resonance modes in [0.6 GHz, 1 GHz] and 7 selected resonance modes in [2.1 GHz, 2.21 GHz] for the 5-cell cavity with 6 H-shaped waveguide ports computed by the proposed solver. Note that the $5^{th}$ mode is the working mode, and all modes above 0.6524 GHz are HOMs. \label{fig:EJ_cavity_5cell_port}}
	\end{figure}

	The cutoff frequencies for the first four propagating modes of the ports are $f_c=$0.776 GHz, 2.132 GHz, 2.132 GHz, and 2.185 GHz. The model discretization leads to $N_J=41,223$ RWG basis functions and ARPACK computes the smallest 100 eigenvalues. Depending on the trial frequency $f$, one can have $0\leq N_M\leq 24$ (i.e., each port supports at most four propagating modes). Each trial frequency $f$ requires about 85 s CPU time \ylrev{when $n_{AK}$=100} and 8.5 GB memory using HODLR on 16 Cori nodes. We perform two frequency searches for [0.6 GHz, 1 GHz] and [2.1 GHz, 2.21 GHz], separately. IERD finds 22 and 35 resonances modes, respectively. Their eigenvalues are plotted in Fig. \ref{fig:cavity_5cell_port_fullspectrum} where each color denotes one eigenmode
	and each dot represents one trial frequency. Again, high-Q modes exhibit narrow bandwidths and smaller minimum eigenvalues, and low-Q modes exhibit higher bandwidths and larger minimum eigenvalues. Fig. \ref{fig:EJ_cavity_5cell_port} shows the surface $|\bm{E}|$ for the first 14 modes in [0.6 GHz, 1 GHz] and 7 selected modes in [2.1 GHz, 2.21 GHz] with their resonance frequencies shown in the subfigure titles. Note that the first 5 modes are below the lowest cutoff frequencies $f_c$=0.776 GHz and the $5^{th}$ mode at 0.65244 GHz is the working mode.   
	
\begin{figure}[!t]
		\centering
		\includegraphics[width=0.9\columnwidth]{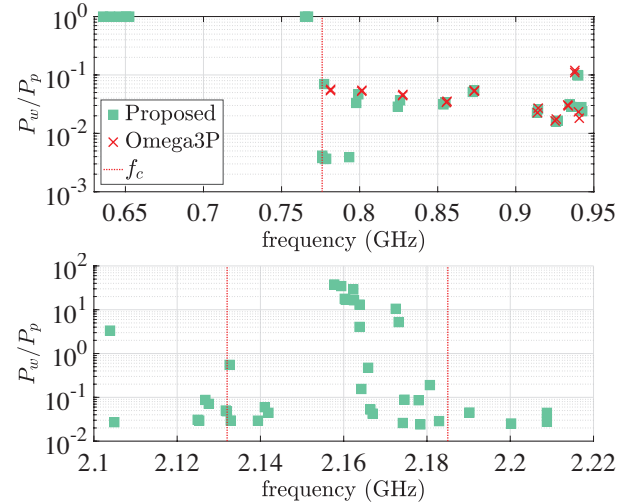}
		\caption{	
			The power ratios $P_w/P_p$ for the 5-cell cavity with 6 H-shaped waveguide ports of all resonance modes in [0.6 GHz, 1 GHz] and [2.1 GHz, 2.21 GHz] detected by the proposed solver, and all resonance modes in [0.776 GHz, 1 GHz] detected by Omega3P. The red vertical lines denote the cutoff frequencies of the ports $f_c=$0.776 GHz, 2.132 GHz, 2.132 GHz, and 2.185 GHz. Below the lowest cutoff frequency 0.776 GHz, $P_p=0$ and we simply set $P_w/P_p=1$. \label{fig:cavity_5cell_ratio}}
	\end{figure}

	Finally, the power ratios $P_w/P_p$ are plotted in Fig. \ref{fig:cavity_5cell_ratio}. Note that CST cannot yet handle waveguide ports not aligned with X/Y/Z plane in Cartesian coordinates and we only compare with Omega3P. We added the eigenmodes detected by Omega3P in [0.776 GHz, 1 GHz] which allows one propagating mode per waveguide port. As can be seen from Fig. \ref{fig:cavity_5cell_ratio}, Omega3P and IERD results match well, but Omega3P is missing a few resonance modes near $f_c=$0.776 GHz.

	\section{Conclusion}\label{sec:con}
	This paper presents IERD, an efficient Integral Equation (IE)-based resonance detection tool for modeling RF cavities with arbitrarily-shaped cross sections and damping waveguide ports supporting multiple propagating modes. IERD leverages fast direct solvers such as HODLR or HODBF from ButterflyPACK, the ARPACK eigensolver to identify resonance modes at specific trial frequencies, and a hybrid Bayesian optimization and downhill simplex algorithm to minimize the number of trial frequencies required. The tool also supports higher-order basis functions and curvilinear surface mesh representation. 
	
	It was shown with numerical experiments that IERD can detect HOMs effectively and accurately over wide frequency bands for realistic RF accelerator cavity modeling with significant lower memory usage compared to FEM-based formulations and tunable CPU requirement depending on the frequency band and allowed maximum numbers of trial frequencies. 
	
	IERD contributes to RF cavity modeling with a novel approach that combines IE methods, fast direct solvers, and advanced optimization algorithms. Its novel hybrid Bayesian optimization and downhill simplex algorithm efficiently locates resonance regions with minimal trial frequencies. Supporting arbitrarily-shaped cross sections, damping waveguide ports, higher-order basis functions, and curvilinear surface mesh, IERD offers a practical and efficient solution for addressing limitations of conventional techniques. Future work includes efficient schemes for post-computing the electric fields inside the cavity from the surface current densities, and integrating IERD into the automatic cavity design process.

	\section*{Acknowledgment}\label{sec:ack}
	This research was supported in part by the Exascale Computing Project (17-SC-20-SC), a collaborative effort of the U.S.~Department of Energy Office of Science and the National Nuclear Security Administration, and in part by the U.S.~Department of Energy, Office of Science, Office of Advanced Scientific Computing Research, Scientific Discovery through Advanced Computing (SciDAC) program through the FASTMath Institute under Contract No.~DE-AC02-05CH11231 at Lawrence Berkeley National Laboratory. This work was supported in part by previous breakthroughs obtained through the Laboratory Directed Research and Development Program of Lawrence Berkeley National Laboratory under U.S. Department of Energy Contract No. DE-AC02-05CH11231. This research used resources of the National Energy Research Scientific Computing Center (NERSC), a U.S.~Department of Energy Office of Science User Facility operated under Contract No.~DE-AC02-05CH11231.

	
	%
	
	

	\ifCLASSOPTIONcaptionsoff
	\newpage
	\fi

	
	
	\bibliographystyle{IEEEtran}
	\bibliography{References.bib}
\end{document}